\documentclass[11pt]{article}
\usepackage{amsmath,amssymb}

\csname @addtoreset\endcsname{equation}{section}

\def\xxx#1           {{\sf hep-th/#1} }

\def\npb#1(#2)#3     {Nucl. Phys. {\bf B#1} (#2) #3 }
\def\rep#1(#2)#3     {Phys. Rept. {\bf #1} (#2) #3 }
\def\pla#1(#2)#3     {Phys. Lett. {\bf A#1} (#2) #3 }
\def\plb#1(#2)#3     {Phys. Lett. {\bf B#1} (#2) #3 }
\def\prl#1(#2)#3     {Phys. Rev. Lett.{\bf #1} (#2) #3 }
\def\prd#1(#2)#3     {Phys. Rev. {\bf D#1} (#2) #3 }
\def\ap#1(#2)#3      {Ann. Phys. {\bf #1} (#2) #3 }
\def\rmp#1(#2)#3     {Rev. Mod. Phys. {\bf #1} (#2) #3 }
\def\cmp#1(#2)#3     {Comm. Math. Phys. {\bf #1} (#2) #3 }
\def\mpla#1(#2)#3    {Mod. Phys. Lett. {\bf A#1} (#2) #3 }
\def\ijmp#1(#2)#3    {Int. J. Mod. Phys. {\bf A#1} (#2) #3 }
\def\cqg#1(#2)#3     {Class. Quant. Grav. {\bf #1} (#2) #3 }
\def\am#1(#2)#3      {Adv. Math. {\bf #1} (#2) #3 }
\def\im#1(#2)#3      {Invent. Math. {\bf #1} (#2) #3 }
\def\jhep#1(#2)#3    {JHEP {\bf #1} (#2) #3 }
\def\npps#1(#2)#3    {Nucl. Phys. Proc. Suppl. {\bf #1} (#2) #3 }
\def\jgp#1(#2)#3     {J. Geom. Phys. {\bf #1} (#2) #3 }
\def\atmp#1(#2)#3    {Adv. Theor. Math. Phys. {\bf #1} (#2) #3}
\def\lmp#1(#2)#3     {Lett. Math. Phys. {\bf #1} (#2) #3}

\begin{document}
\input feynman
\thispagestyle{empty}
\null\vskip-24pt \hfill CERN-TH/2001-285 \vskip-10pt \hfill LAPTH-872/01

\begin{center}
\vskip 0.2truecm {\Large\bf
Universal properties of superconformal OPEs for 1/2 BPS operators in
$\mathbf{3\leq {D}\leq 6}$\  \footnote{To be published in NJP Focus Issue: {\it
Supersymmetry in condensed matter and high energy physics}} \vskip 0.5truecm }

{\Large Sergio Ferrara\footnote{Laboratori Nazionali di Frascati, INFN, Italy,
and Department of Physics and Astronomy, University of California, Los Angeles,
CA 90095, USA} and Emery Sokatchev\footnote{On leave of absence from
Laboratoire d'Annecy-le-Vieux de Physique Th{\'e}orique  LAPTH, B.P. 110,
F-74941 Annecy-le-Vieux et l'Universit{\'e} de Savoie}} \vskip 0.4truecm

\vskip 0.3cm {CERN Theoretical Division, CH 1211 Geneva 23, Switzerland}
\end{center}

\vskip 1truecm \Large
\centerline{\bf Abstract} \normalsize We give a general analysis of OPEs of 1/2
BPS superfield operators for the $D=3,4,5,6$ superconformal algebras OSp(8/4,R),
PSU(2,2), F${}_4$ and OSp($8^*/4$) which underlie maximal AdS supergravity in
$4\leq
D+1\leq 7$. \\
The corresponding three-point functions can be formally factorized in a way
similar to the decomposition of a generic superconformal UIR into a product of
supersingletons. This allows for a simple derivation of branching rules for
primary superfields. The operators of protected conformal dimension which may
appear in the OPE are classified and are shown to be either 1/2 or 1/4 BPS, or
semishort. As an application, we discuss the ``non-renormalization" of extremal
$n$-point correlators.

\newpage
\setcounter{page}{1}\setcounter{footnote}{0}

\section{Introduction}\label{intro}

Maximal AdS supergravities in  4,5,6 and 7 dimensions are dual to
superconformal field theories on the world volume of $M_2$, $D_3$, $D_4/D_8$ and
$M_5$ branes, respectively \cite{AGMOO}. Although only the $D_3$ brane dynamics has a
perturbative description in the superconformal regime, some general properties
of abstract superconformal field theories can be obtained by using the BPS
nature of a certain class of superconformal primary operators and the model
independent nature of OPEs (for reviews see, e.g., \cite{Frrev,Brev,HWrev}).

Superconformal algebras satisfying the Haag-Lopuszanski-Sohnius theorem \cite{Haag:1975qh} exist only for $D\leq 6$ \cite{Nahm:1978tg}. The maximal ones for $D=3,4,5$ and 6 are   OSp(8/4,R),
PSU(2,2), F${}_4$ and OSp($8^*/4$),  respectively.

In the classification of UIRs of superconformal algebras an important r\^ole is
played by the representations with ``quantized" conformal dimension since in
the quantum field theory framework they correspond to operators with
``protected" scaling dimension and therefore imply ``non-renormalization
theorems" at the quantum level.

There are different classes of operators with protected dimension in
supersymmetric field theories. One of them are the BPS operators corresponding
to different fractions of preserved supersymmetry. In the superfield language
these operators are the natural generalization of the ``chiral" superfields of
$N=1$, $D=4$ SUSY theories \cite{AFSZ,FS1}. For $N$-extended theories,
independence of a certain subset of the Grassmann coordinates in superspace
corresponds to a certain fraction of preserved supersymmetry. These operators,
like $N=1$ chiral superfields, form a ring under multiplication. Their natural
description is ``harmonic superspace" \cite{GIK1,TheBook} and the corresponding
BPS superfields are called Grassmann analytic \cite{GIO}. A well-known example
\cite{hw1,Ferrara:1998ej,AF} of 1/2 BPS operators (which depend only on half of the $\theta$s)
in $N=4$, $D=4$ super-Yang-Mills theory are the operators corresponding to
Kaluza-Klein excitations of type IIB SUGRA on AdS${}_5\times S^5$
\cite{RomNieu,gm}.

However, it is known from explicit calculations in $N=4$, $D=4$ SYM and in
$D=5$ AdS SUGRA \cite{AFP1,AFP2,AEPS,AES,BKRSkonishi} that there exist other
``protected operators" which do not correspond to any BPS states. They form
another class of operators with protected dimension called ``current-like" or
``semishort" \cite{FrGuWa,FS1}. The corresponding superfields satisfy
differential constraints in superspace which imply that certain terms in the
$\theta$ expansion are missing (but not entire $\theta$s, as in the BPS case).
Only in very particular cases these constraints affect the space-time
dependence and the superfields contain true space-time currents.

The protected supermultiplet found in \cite{AFP1,AFP2,AEPS,BKRSkonishi} was identified as a  semishort multiplet in \cite{ES,HH}. Its lowest component is a scalar of conformal dimension 4 in the $\mathbf{20}$ of
SU(4). Other examples of semishort superfields are certain K-K
excitations of type IIB SUGRA on AdS${}_5\times T_{11}$ \cite{Rom} studied in
\cite{CerFer}. In fact, the K-K excitations of the graviton, while being 1/2
BPS states in the $N=8$, $D=5$ AdS supergravity, are semishort multiplets in
the dual $N=1$, $D=4$ superconformal field theory formulated in \cite{KleWit}.

In the present paper we give a unified discussion of all superconformal field
theories with maximal supersymmetry in $3\leq D\leq 6$, dual to AdS maximal
supergravities in $4\leq D+1\leq 7$. We study the general branching rules (or
equivalently, the OPE) of two 1/2 BPS states into a third, a priori arbitrary
state. To this end we examine the three-point functions that two 1/2 BPS
operators can form with any other operator. Such three-point functions are
uniquely determined by conformal supersymmetry. In addition, imposing the
conditions of BPS shortness at two of the points leads to selection rules on
the operator at the third point. In this way we find out, in all different
channels specified by the R symmetry quantum numbers, which are the allowed types of
operators that can appear in such OPEs.

This analysis clarifies the possibility to have operators with ``anomalous
dimension" in certain channels (such as the Konishi multiplet) and the
occurrence of only operators with ``protected dimension" in other channels.
These results find applications, for example, in the proof of the
non-renormalization of the so-called ``extremal" higher $n$-point functions of
1/2 BPS operators \cite{DHoFrMaMaRa,DP}.

The method we propose here differs from earlier approaches to the same problem in $D=4$ \cite{AES,ES,HH,HH2} and in $D=6$ \cite{EFS} in the sense that it avoids studying the detailed structure of
the superspace three-point functions. Instead, we put the emphasis on the
purely group-theoretic aspects of the problem. We efficiently exploit a newly
established formal factorization property of the three-point functions. It reflects
the possibility \cite{AFSZ,Ferrara:2000zg,FS2,FS1,HHowe} to realize the generic superconformal
UIRs as composite operators made out of one basic constituent, the so-called
``supersingleton" (or massless supermultiplet)
\cite{Fron,Duf,Nicolai:1984gb,Gunaydin:1985wc}. This description of the
three-point functions results in a unified treatment of all theories where the
supersingleton constituents are identified with the microscopic ``degrees of
freedom" of the brane world volume dynamics.\footnote{The basic degrees of freedom of the brane world volume actually carry a color index $N_c$ \cite{AGMOO} which, however, is not relevant to our discussion. Details about the realization of some of the operators considered here as gauge invariant  composites in $N=4$, $D=4$ SYM can be found in \cite{BKRSkonishi,HH}.}

We show that the selection rules or, to put it differently, the ``protection
mechanism" for certain channels in the OPE of two 1/2 BPS operators has a very
simple origin which can be illustrated by the following example from ordinary
conformal field theory \cite{FT,FP}. Consider the three-point function of two
scalar fields with a rank $s$ symmetric tensor field:
\begin{equation}\label{counterexa}
  \langle \phi_A(1) \phi_B(2) j^{\{\mu_1\cdots\mu_s\}}(3) \rangle \;.
\end{equation}
These fields have conformal dimensions $\ell_A,\ell_B$ and $\ell$,
respectively. Now, suppose that the scalars are massless (``singletons"),
\begin{equation}\label{singlett}
  \square\phi_A=\square\phi_B=0\,.
\end{equation}
These equations are conformally invariant only if the scalars have the
canonical dimension $\ell_A=\ell_B=(D-2)/2$. But this is not all: A direct
calculation shows that the condition (\ref{singlett}) also fixes the dimension
of the tensor $j^{\{\mu_1\cdots\mu_s\}}$ at its canonical value $\ell=s+D-2$
and, moreover, {\it forces this tensor to be conserved}. Thus, imposing a
condition on the operators at points 1 and 2 of the three-point function can
have the effect of ``protecting" the operator at point 3.

A similar phenomenon takes place with the three-point functions involving two
1/2 BPS short multiplets. We write them down formally in a factorized form in which the
R symmetry quantum numbers at the third point are associated with a BPS factor
whereas the spin and the (possibly anomalous) dimension are carried by a
singlet factor. Depending on the choice of the R symmetry irrep at point 3,
this singlet factor turns out to be either trivial, or of the type
(\ref{counterexa}), (\ref{singlett}) above, or unconstrained. Correspondingly,
we find the following selection rules for the third operator: It is either BPS
short, or semishort, or unconstrained. Thus, in the first two cases the
operator at point 3 is ``protected" while in the third case it is
``unprotected".

The paper is organized as follows. In Section 2 we describe the supersingleton
degrees of freedom (brane supercoordinates) for all these theories and their
relation with supermultiplet shortening. We also explain how various
superconformal UIRs, including different kinds of BPS short and semishort
multiplets, can be obtained as composite operators made out of supersingletons.
In Section 3 we give a unified treatment of the three-point functions involving
two 1/2 BPS operators by writing them down in the factorized form described
above and deriving the selection rules for the third operator. The results are
interpreted in Section 4 where particular attention is paid to the occurrence
and the r\^ole of the semishort protected multiplets. In Section 5 we give an
application of these results in the form of a general non-renormalization
theorem for ``extremal" $n$-point correlators of 1/2 BPS multiplets.

\section{Supersingletons}\label{singl}

\subsection{Standard description of the supersingletons}

Supersingletons are massless representations of the D-dimensional
superconformal algebra or equivalently, superfields satisfying conformally
covariant massless field equations. It is well known \cite{Siegel2,EHWa,FeFo}
 that there exist only a finite number
of them in $D=3,5$ while there are infinite sets in $D=4,6$. Here we restrict
ourselves to the so-called 1/2 BPS supersingletons which have been identified
with the basic brane degrees of freedom in the context of the AdS/CFT
correspondence. These supersingletons are ``ultrashort" supermultiplets with
maximal spin $s_{max}=1/2$ in $D=3,5$ and $s_{max}=1$ in $D=4,6$. Such
supermultiplets can be characterized by the quantum numbers of their lowest
component: vanishing Lorentz spin, canonical conformal dimension of a massless
scalar $(D-2)/2$ and R symmetry irrep with Dynkin labels (DL) according to the
following list:
\begin{equation}\label{irreps}
    \begin{array}{lll}
    D=3: & [0010] & \mbox{$\mathbf{8_s}$ of SO(8)} \\
    D=4: & [010] & \mbox{{\bf 6} of SU(4)} \\
    D=5: & [1] & \mbox{{\bf 2} of SU(2)} \\
    D=6: & [01] & \mbox{{\bf 5} of USp(4)}
  \end{array}
\end{equation}
Note that in the case $D=3$ there are two inequivalent choices of the basic
supersingleton, e.g., $[0010]$ ($\mathbf{8_s}$) and $[0001]$ ($\mathbf{8_c}$),
and a third one, $[0100]$ ($\mathbf{8_v}$), related to the first two by SO(8)
triality. Here we restrict ourselves to the OPE of two supersingletons of the
same type $\mathbf{8_s}$.

These supersingletons can be described in terms of scalar superfields carrying
external R symmetry indices according to (\ref{irreps}) and satisfying the
following on-shell constraints \cite{Siegel,HST,Howe:1983fr,Parkkk,Howe}:
\begin{equation}\label{onshellcon}
    \begin{array}{ll}
    D=3: & D^i_\alpha\; W_a = \frac{1}{8}(\Gamma^i\tilde\Gamma^j)_{ab} D^j_\alpha\; W_b \\
    D=4: & D^{(k}_\alpha\; W^{[i)j]} = 0\;, \quad \bar D_{\dot\alpha \{k}\; W^{[i\}j]} = 0\\
    D=5: & D^{(k}_\alpha\; W^{i)} = 0 \\
    D=6: & D^{(k}_\alpha\; W^{[i)j]} = 0
  \end{array}
\end{equation}
Here the indices $i,j,k$ belong to the fundamental representation (or its
complex conjugate) of the R symmetry groups, except for the case $D=3$, where
$i$ is an $\mathbf{8_v}$ index and $a,b$ are $\mathbf{8_s}$ indices; $\Gamma^i$
denote the gamma matrices of SO(8). The symbols $()$, $[]$ and $\{\}$ mean
symmetrization, antisymmetrization and traceless part, respectively. These
constraints eliminate most of the components of the superfields and put the
remaining ones on the massless shell.


\subsection{Supersingletons as 1/2 BPS short superfields}

Harmonic superspace allows us to write down all these supersingletons as 1/2
BPS short (or Grassmann analytic) superfields depending only on half of the
Grassmann variables. To this end we introduce harmonic variables $u$ on the
coset $R/H$ where $R$ is the R symmetry group and $H=[{\rm
U}(1)]^{\mbox{\scriptsize  rank}(R)}$ is its maximal torus. With their help we
can covariantly project $R$ indices onto $H$ ones. Then suitable projections of
the original massless superfields become Grassmann analytic objects, i.e.,
superfields which depend only on half of the full set of Grassmann coordinates.
The most convenient way is to always choose the projection of the external R
indices onto what corresponds to the highest weight state (HWS) of the
representation. The details can be found in \cite{FS1} but they are not
essential for the argument we present here. Let us just list these Grassmann
analytic superfields:
\begin{equation}\label{supersingl}
  \begin{array}{ll}
    D=3: & W^{+(+)[+]}(x,\theta^{++}, \theta^{(++)}, \theta^{[+]\{\pm\}},u) \\
    D=4: & W^{12}(x,\theta_3,\theta_4,\bar\theta^1,\bar\theta^2,u) \\
    D=5: & W^1(x,\theta^1,u) \\
    D=6: & W^{12}(x,\theta^1,\theta^2,u)
  \end{array}
\end{equation}
In the case $D=3$ the four sets of U(1) charges are denoted by
$\pm(\pm)[\pm]\{\pm\}$; in the other cases it is more convenient to use the
individual projections of the indices of the fundamental representation
($i=1,2$ for SU(2) and $i=1,2,3,4$ for SU(4) and USp(4)) to label the different
states of an R symmetry irrep.

These superfields are in general functions of the harmonic variables having
infinite expansions on the harmonic coset $R/H$. The condition which cuts
these expansions down to polynomials in the harmonics $u$ is the condition of
{\it R symmetry irreducibility}. It takes the familiar form of the definition
of a HWS:
\begin{equation}\label{HWS}
  {D^\uparrow_u} W=0\;.
\end{equation}
Here ${D^\uparrow_u}$ denotes the set of raising (step-up or creation)
operators of the group $R$ realized in the form of covariant harmonic
derivatives on the coset $R/H$. If one uses a complex parametrization of the
coset, conditions (\ref{HWS}) become covariant Cauchy-Riemann conditions
(harmonic analyticity \cite{hh}). The combination of Grassmann analyticity
(\ref{supersingl}) with irreducibility under the R symmetry group (\ref{HWS})
is equivalent to the original formulation (\ref{onshellcon}) of the
supersingletons.

Supersingletons are ``ultrashort" superfields in the sense that their $\theta$
expansion contains just a few massless fields. Here we show only the bosonic
content of the G- and H-analytic superfields (\ref{supersingl}):
\begin{equation}\label{comp}
  \begin{array}{ll}
    D=3: & W^{+(+)[+]} = \phi_a(x) u^{+(+)[+]}_a + \mbox{derivative terms} \\
    D=4: & W^{12} = \phi^{[ij]}(x) u^1_i u^2_j +
     (\theta_3 \sigma^{\mu\nu}\theta_4 + \bar\theta^1\sigma^{\mu\nu}\bar\theta^2)
      F_{\mu\nu}(x) +\mbox{d.t.} \\
    D=5: & W^1 = \phi^i(x) u^1_i + \mbox{d.t.}\\
    D=6: & W^{12} = \phi^{[ij]}(x) u^1_i u^2_j +
     \theta^1 \gamma^{\mu\nu\lambda}\theta^2 F_{\mu\nu\lambda}(x) +\mbox{d.t.}
  \end{array}
\end{equation}
The massless scalar fields $\phi$ belong to the R symmetry irreps listed in
(\ref{irreps}), the on-shell two-form and three-form field strengths $F$ are
singlets.

Concluding this subsection we mention that (\ref{supersingl}) is not the only
possible realization of the supersingletons as Grassmann analytic superfields.
Instead of projecting the external R symmetry indices onto the HWS, we could
take any other state and accordingly choose the half of $\theta$'s that the
superfield depends on, e.g.,
\begin{equation}\label{altern}
  \begin{array}{ll}
    D=3: & W^{+(+)[-]}(x,\theta^{++}, \theta^{(++)}, \theta^{[-]\{\pm\}},u) \\
    D=4: & W^{13}(x,\theta_2,\theta_4,\bar\theta^1,\bar\theta^3,u) \\
    D=6: & W^{13}(x,\theta^1,\theta^3,u)
  \end{array}
\end{equation}
(there is no need to do this in the case $D=5$). Unlike the superfields
(\ref{supersingl}), the new ones are not harmonic analytic (since they are not
HWS), i.e., they are not annihilated by all of the raising operators (harmonic
derivatives). Instead, they are related to the HWS (\ref{supersingl}) by the
action of some of the raising operators:
\begin{equation}\label{stepup}
  D^{[++]} W^{+(+)[-]} = W^{+(+)[+]}\;, \qquad D^2_3 W^{13} = W^{12}\;.
\end{equation}
The use of this alternative realization is explained in the next subsection.

\subsection{BPS short multiplets as products of supersingletons}
\label{compoBPS}

An important advantage of the description of the supersingletons as Grassmann
and harmonic analytic superfields is the possibility to obtain new BPS short
objects by simply multiplying the basic supersingletons \cite{FS1}. The reason is that
analytic objects form a ring structure, i.e., a set closed under
multiplication.

Thus, any power $[W]^k$ is automatically Grassmann analytic, i.e., depends on
the same half of the Grassmann variables, recall (\ref{supersingl}). Further,
since the supersingleton $W$ is the HWS of the R symmetry irrep with DL listed
in (\ref{irreps}), the power $[W]^k$ satisfies the constraint which defines it
as the HWS of one of the following irreps:
\begin{equation}\label{1/2BPSm}
     \begin{array}{lcl}
    D=3:  & [00k0] &  \mbox{of SO(8)} \\
    D=4:  & [0k0] & \mbox{of SU(4)} \\
    D=5:  & [k] & \mbox{of SU(2)} \\
    D=6:  & [0k] & \mbox{of USp(4)}
  \end{array}
\end{equation}

This way of obtaining the 1/2 BPS operators as composite objects makes clear
the important fact that the implications of the BPS shortness conditions depend
on the quantum numbers of the superfield. Consider the 1/2 BPS short
superfields
\begin{equation}\label{kssgl'}
\mbox{BPS}_{1/2}^{(k)} : \quad \left\{
  \begin{array}{llll}
    D=3: & {\cal W}^{[00k0]}(x,\theta^{++}, \theta^{(++)}, \theta^{[+]\{\pm\}},u)
     &\Leftrightarrow& [W^{+(+)[+]}]^k \\
    D=4: & {\cal W}^{[0k0]}(x,\theta_3,\theta_4,\bar\theta^1,\bar\theta^2,u)
     & \Leftrightarrow& [W^{12}]^k \\
    D=5: & {\cal W}^{[k]}(x,\theta^1,u)
     &\Leftrightarrow&  [W^{1}]^k  \\
    D=6: & {\cal W}^{[0k]}(x,\theta^1,\theta^2,u)
     &\Leftrightarrow& [W^{12}]^k
  \end{array} \right.
\end{equation}
where in the left column we have indicated the R symmetry DL instead of the
U(1) charges. These superfields satisfy the same conditions of BPS shortness
(i.e., of Grassmann and harmonic analyticity), but their component content
strongly depends on the value of $k$. In the case of the supersingleton ($k=1$)
we have seen that the combination of the two conditions puts the superfield on
the massless shell. An even stronger constraint is obtained for $k=0$: A
singlet analytic object can only be a constant, as follows from the obvious
property $(W)^0=1$. However, for $k\geq 2$ the constraints become much weaker.
In particular, the first component of the superfield, a scalar of dimension
$k(D-2)/2$, satisfies no constraint whatsoever.\footnote{It should be mentioned
that for $k=2$ in $D=4,5,6$ and for $k=2,3$ in $D=3$ some of the higher
components of this composite superfield are conserved vectors or tensors. The
best known example is the $N=4$, $D=4$ SYM stress-tensor multiplet which is
described by the supersingleton bilinear $W^{12}W^{12}$.} Indeed, if we realize
the 1/2 BPS superfield as a composite operator $(W)^k$ for $k\geq 2$, we see
that the first component is a generic scalar composite made out of the massless
scalars $\phi$ from (\ref{comp}). This crucial distinction among the cases
$k=0$, $k=1$ and $k\geq 2$ is at the origin of the selection rules for the
three-point functions which are derived in Section \ref{selrul}.

Another possibility of obtaining BPS short composite operators is to multiply
together two different realizations of the basic supersingleton. For instance,
the product of Grassmann analytic superfields of the types (\ref{supersingl})
and (\ref{altern}), or of any of their powers, is a superfield which does not
depend on 1/4 of the full set of $\theta$'s. According to the AdS terminology,
such operators are called ``1/4 BPS short":
\begin{equation}\label{kssgl}
{\mbox{BPS}_{1/4}^{(jp)}} : \quad \left\{
     \begin{array}{llll}
    D=3: & {\cal W}^{[0jp0]}(x,\theta^{++}, \theta^{(++)},
    \theta^{[\pm]\{\pm\}},u)
     &\Leftrightarrow& [W^{+(+)[+]}]^{p+j}[W^{+(+)[-]}]^j\\
    D=4: & {\cal W}^{[jpj]}(x,\theta_2,\theta_3,\theta_4,
    \bar\theta^1,\bar\theta^2,\bar\theta^3,u)
     & \Leftrightarrow& [W^{12}]^{p+j}[W^{13}]^j \\
    D=6: & {\cal W}^{[2j,p]}(x,\theta^1,\theta^2,\theta^3,u)
     &\Leftrightarrow& [W^{12}]^{p+j}[W^{13}]^j
  \end{array} \right.
\end{equation}
This time, since the factor of the type (\ref{altern}) is not harmonic
analytic, the product does not automatically define a HWS of a new R symmetry
irrep. This can be achieved by imposing further irreducibility conditions. For
example, in the case $D=3$ the harmonic condition (recall (\ref{stepup}))
\begin{equation}\label{D3cond}
  D^{[++]}\left([W^{+(+)[+]}]^{p+j}[W^{+(+)[-]}]^j\right) = 0
\end{equation}
defines the HWS of the irrep $[0jp0]$ of SO(8). Similarly, in the case $D=4$
(or $D=6$) the condition
\begin{equation}\label{D4cond}
  D^2_3\left([W^{12}]^{p+j}[W^{13}]^j\right) = 0
\end{equation}
turns the product into the HWS of the irrep $[jpj]$ of SU(4) (or $[2j,p]$ of
USp(4)).

We remark that eq. (\ref{kssgl}) does not exhaust the list of composite BPS
objects obtained by multiplying various realizations of the basic
supersingleton \cite{FS1}. For example, in $D=3$ one could have 1/8 and 3/8, in
$D=4$ 1/8 BPS multiplets, etc. We do not consider them here because they are
associated with R symmetry irreps different form those appearing in the OPE of
two 1/2 BPS operators (see eq. (\ref{tensprod})).

\subsection{Semishort multiplets} \label{Semishort multiplets}

In what follows the so-called ``semishort" (or ``current-like") multiplets will
play an important r\^ole. In this section we give a brief summary of the origin
of such multiplets as limiting cases or as isolated points in the series of
UIRs of the superconformal algebras. We also give their realization as
composite operators made out of supersingletons which satisfy some
``current-like" superspace constraints.\footnote{Note that if the supersingletons carry a color index $N_c$, under which the composite is a singlet, there are in principle different operators with the same spin and R symmetry quantum numbers \cite{AF,Skiba:1999im,BKRSkonishi,HH}.}

The semishort multiplets are to some extent the analogs of the ``conserved"
tensor representations of the ordinary conformal group SO($D$,2). It is well
known that the rank $s$ symmetric traceless tensor field
$j^{\{\mu_1\cdots\mu_s\}}(x)$ of the so-called ``canonical" dimension
$\ell=s+D-2$   forms a reducible but indecomposable representation of the
conformal group SO($D$,2) \cite{FP}. This means  that its divergence
$\partial_{\mu_1}j^{\{\mu_1\cdots\mu_s\}}$ transforms covariantly and can be
set to zero. The resulting ``transverse" tensor is already
irreducible.\footnote{The representation is indecomposable because the
``longitudinal" part cannot be projected out by a local conformal operator.}

The conserved tensors can be viewed as a limiting case of the generic series of
UIRs of SO($D$,2).  A generic conformal operator carrying conformal dimension
$\ell$ and Lorentz spin $s$ can be written down in the following composite
form:
\begin{equation}\label{generconf}
  {\cal O}_{\ell s} =  j^{\{\mu_1\cdots\mu_s\}}\;\phi^\delta \;.
\end{equation}
Here $j^{\{\mu_1\cdots\mu_s\}}$ is a conserved tensor and $\phi$ is a massless
scalar (``singleton") field:
\begin{equation}\label{single}
  \partial_{\mu_1}j^{\{\mu_1\cdots\mu_s\}}=0,\quad \ell_j=s+D-2\;; \qquad
  \square\phi=0, \quad \ell_\phi= (D-2)/2\;.
\end{equation}
Note that  $j^{\{\mu_1\cdots\mu_s\}}$ can itself be represented as a composite
operator made out of singletons, e.g., for $s=1$
\begin{equation}\label{curr}
  j^\mu = i(\phi\partial^\mu\phi' -  \phi'\partial^\mu\phi)
\end{equation}
where $\phi'$ is another copy of the singleton.  The parameter $\delta$ in eq.
(\ref{generconf}) can take non-integer values, $\delta\geq 0$ (for $s>0$) or
$\delta\geq1$ (for $s=0$). This accounts for the possible ``anomalous"
dimension of the operator ${\cal O}_{\ell s}$ subject to the unitarity bound
$\ell\geq s+D-2$ (for $s>0$) or $\ell\geq (D-2)/2$ (for $s=0$). From the
``composite" form (\ref{generconf}) it is clear that the unitarity bound is
saturated if $\delta=0$, $s>0$ (no massless scalar appears) or if $\delta=1$
and  $s=0$. Thus, the conserved tensor is at the threshold of the continuous
series of UIRs represented by the composite operators (\ref{generconf}).

A similar phenomenon takes place in the classification of the superconformal
UIRs. Let us first recall some of the known series of UIRs \cite{FF,dp,Minw2}. We restrict ourselves to those which can possibly form a three-point function with two 1/2 BPS UIRs. They
must carry Lorentz indices corresponding to a symmetric traceless tensor of
rank $s$ and R symmetry quantum numbers which are listed in (\ref{tensprod}).

{\bf OSp(8/4)} ($D=3$): The Lorentz quantum number (spin) is an integer $J=s$
and we are dealing with SO(8) representations of the type $[0a_2a_30]$. There
exist two series of UIRs:
\begin{eqnarray}
   \mbox{A)}&&\ell \geq 1+s + a_2 + \frac{1}{2}a_3\,, \nonumber\\
   \mbox{B)}&&s=0, \quad \ell = a_2 + \frac{1}{2}a_3\,.  \label{3A}
\end{eqnarray}
The discrete series B contains the BPS multiplets.

{\bf PSU(2,2/4)} ($D=4$): We consider Lorentz spins $J_1=J_2=s/2$ and SU(4)
representations of the type $[a_1a_2a_1]$. Two of the three existing series of
UIRs are relevant in this case:
\begin{eqnarray}
   \mbox{A)}&& \ell \geq 2 + s + 2a_1 + a_2\,, \nonumber\\
   \mbox{C)}&&s=0, \quad \ell = 2a_1 + a_2\,.  \label{4A}
\end{eqnarray}
The discrete series C contains the BPS multiplets.

{\bf F${}_4$} ($D=5$): We consider Lorentz spins $J_1=0$, $J_2=s$ and  SU(2)
representations $[a_1]$.  There exist three series of UIRs:
\begin{eqnarray}
   \mbox{A)}&& \ell \geq 4 + s + \frac{3}{2}a_1\,, \nonumber\\
   \mbox{B)}&&  \ell = 3 + s + \frac{3}{2}a_1\,, \nonumber\\
   \mbox{C)}&&s=0, \quad \ell = \frac{3}{2}a_1\,.  \label{5A}
\end{eqnarray}
Series B is an ``isolated" series and series C contains the BPS multiplets.

{\bf OSp(8${}^*$/4)} ($D=6$): We consider Lorentz spins $J_1=0$, $J_2=s$,
$J_3=0$ and USp(4) representations of the type $[a_1a_2]$ (with even $a_1$).
There exist four series of UIRs:
\begin{eqnarray}
   \mbox{A)}&& \ell \geq 6+s + a_2 + 2(a_1+a_2)\,, \nonumber\\
   \mbox{B)}&&  \ell = 4 + s + 2(a_1+a_2)\,, \nonumber\\
   \mbox{C)}&&  s=0, \quad \ell = 2 + 2(a_1+a_2)\,, \nonumber\\
   \mbox{D)}&&  s=0, \quad \ell = 2(a_1+a_2)\,.  \label{6A}
\end{eqnarray}
Series B and C are isolated and series D contains the BPS multiplets.

In close analogy with the factorization (\ref{generconf})  of the conformal
UIRs in terms of singletons, we can write down an operator  ${\cal O}_{\ell
s}^{[a_i]}$ belonging to a generic superconformal UIR labeled by its conformal
dimension $\ell$, spin $s$ and R symmetry DL $a_i$ as a formal product of three
factors \cite{FS1}:
\begin{equation}\label{genersuperconf}
  {\cal O}_{\ell s}^{[a_i]} = J^{\{\mu_1\cdots\mu_s\}}\; \Phi^\delta\;
  \mbox{BPS}^{[a_i]}\;.
\end{equation}
The first factor accounts for the Lorentz spin, the second for the conformal
dimension and the third for the R symmetry labels of the composite operator
${\cal O}_{\ell s}^{[a_i]}$.

Each of these factors can in turn be viewed as a ``fake composite" operator obtained
from the basic supersingletons. Thus, the spin factor has the form of a
bilinear composite of dimension $s+D-2$, e.g., for $s=1$
\begin{equation}\label{supercur}
    \begin{array}{ll}
    D=3: & J^\mu = (\gamma^\mu)^{\alpha\beta} D^i_\alpha W_a(\Gamma^i\tilde\Gamma^j)_{ab}
    D^j_\beta W'_b + 32i(W_a\partial^\mu W'_a - W'_a\partial^\mu W_a) \\
    D=4: & J^\mu = (\sigma^\mu)^{\alpha\dot\alpha} D^i_\alpha W'_{ik}
    \bar D_{\dot\alpha j} W^{jk} + i (W^{ij}\partial^\mu W'_{ij} - W'_{ij} \partial^\mu W^{ij}) \\
    D=5: & J^\mu = (\gamma^\mu)^{\alpha\beta} D^i_\alpha W_i D^j_\beta W_j + i W^i\partial^\mu W_i \\
    D=6: & J^\mu = (\gamma^\mu)^{\alpha\beta} D^i_\alpha W_{ik}\Omega^{kl}
    D^j_\beta W'_{jl} + i (W^{ij}\partial^\mu W'_{ij} - W'_{ij} \partial^\mu W^{ij})
  \end{array}
\end{equation}
and similarly for $s>1$  (see, e.g., \cite{HST}). Note that these composites
satisfy superspace ``conservation conditions" following from the massless
superfield equations (\ref{onshellcon}). Using spinor notation, they can be
written down as follows:
\begin{equation}\label{concon}
  \begin{array}{ll}
    D=3: & D^{i\alpha_1} J_{\alpha_1\cdots\alpha_s} =0  \\
    D=4: & D^{i\alpha_1} J_{\alpha_1\cdots\alpha_s}^{\dot\alpha_1\cdots\dot\alpha_s} = 0, \quad
           \bar D_{i\dot\alpha_1}  J_{\alpha_1\cdots\alpha_s}^{\dot\alpha_1\cdots\dot\alpha_s} = 0 \\
    D=5,6: & \epsilon^{\delta\gamma\alpha_1\beta_1}D^i_\gamma J_{\alpha_1\cdots\alpha_s\;
    \beta_1\cdots\beta_s}=0
  \end{array}
\end{equation}
These superspace constraints imply space-time conservation conditions on
certain components of $J^{\{\mu_1\cdots\mu_s\}}$, including the lowest
component $j^{\{\mu_1\cdots\mu_s\}} = J^{\{\mu_1\cdots\mu_s\}}(\theta=0)$ which
has canonical dimension $s+D-2$. For this reason we call these composites
``supercurrents".

Next, the scalar factor $\Phi$ in eq. (\ref{genersuperconf}) is another
bilinear composite made out of the basic supersingleton, $W_aW_a$ ($D=3$),
$\epsilon^{ijkl}W_{ij}W_{kl}$ ($D=4,6$) and $\epsilon^{ij} W_i W'_j$ ($D=5$).
It also satisfies a superspace conservation constraint whose explicit form we
do need here. As a result,  the ``supercurrent" $\Phi$ contains conserved
tensors, including a vector at the level $\theta\gamma_\mu\theta$. The power
$\delta$ of this scalar factor accounts for the possible anomalous dimension of
the composite operator ${\cal O}_{\ell s}^{[a_i]}$. By choosing the appropriate
values of $\delta$ we can reproduce both the continuous and isolated series of
UIRs (setting $s=\delta=0$ gives the BPS series).

Finally, the BPS factor in eq. (\ref{genersuperconf}) is made out of the
Grassmann analytic supersingletons as explained in Section \ref{compoBPS}.

Now, the formal factorization (\ref{genersuperconf}) allows us to explain the
origin of the semishort multiplets as limiting cases of the generic series of
UIRs. The idea is to keep just one ``supercurrent" factor in
(\ref{genersuperconf}) as well as the BPS factor. Thus, we either set $s>0$ and
$\delta=0$:
\begin{equation}\label{semispin}
  {\cal S}^{\{\mu_1\cdots\mu_s\}\;[a_i]} =  J^{\{\mu_1\cdots\mu_s\}}\;
  \mbox{BPS}^{[a_i]}
\end{equation}
or $s=0$ and $\delta=1$:
\begin{equation}\label{seminospin}
  {\cal S}^{[a_i]} =  \Phi\;
  \mbox{BPS}^{[a_i]}\;.
\end{equation}
In both cases the conformal dimension of ${\cal S}$ is ``quantized" (fixed):
\begin{equation}\label{specialell}
    \begin{array}{ll}
    D=3: & \ell = s + 1 + a_2 + \frac{1}{2} a_3 \\
    D=4: & \ell = s + 2 + 2a_1+a_2  \\
    D=5: & \ell = s + 3 + \frac{3}{2}a_1 \\
    D=6: & \ell = s + 4 + 2(a_1+a_2)
  \end{array}
\end{equation}
According to the classification of UIRs given above, these values correspond to
the saturated unitarity bound of the continuous series A for $D=3,4$ or to the
isolated series B for $D=5,6$.

The defining property of the semishort superfields is that they satisfy some
superspace constraints obtained as the intersection of the supercurrent
constraints (\ref{concon}) (or of their analogs for the scalar supercurrent
$\Phi$) and the Grassmann analyticity constraints on the BPS factor. For
example, in the case of 1/4 BPS shortening (see (\ref{kssgl})) only the
following projections of eqs. (\ref{concon}) hold:
\begin{equation}\label{conconproj}
  \begin{array}{ll}
    D=3: & D^{++\alpha_1} {\cal S}_{\alpha_1\cdots\alpha_s} = D^{(++)\alpha_1} {\cal S}_{\alpha_1\cdots\alpha_s} =0  \\
    D=4: & D^{1\alpha_1} {\cal S}_{\alpha_1\cdots\alpha_s}^{\dot\alpha_1\cdots\dot\alpha_s} = 0, \quad
           \bar D_{4\dot\alpha_1}  {\cal S}_{\alpha_1\cdots\alpha_s}^{\dot\alpha_1\cdots\dot\alpha_s} = 0 \\
    D=6: & \epsilon^{\delta\gamma\alpha_1\beta_1}D^4_\gamma {\cal S}_{\alpha_1\cdots\alpha_s\;
    \beta_1\cdots\beta_s}=0
  \end{array}
\end{equation}
These constraints are significantly weaker in the sense that the corresponding
``current-like" superfield does not contain any conserved tensor components.
Without going into the details of the $\theta$ expansion, this is quite clear
from the factorized form of the ``current-like" operators which is at least
trilinear in the basic supersingletons. The r\^ole of the constraints now is to
simply eliminate some components in the $\theta$ expansion (but not entire
projections of $\theta$s, as in the BPS case). Thus, the  ``current-like"
multiplets do not reach the maximal spin of the generic superfield of the same
type, and for this reason we call them ``semishort".

\section{Selection
rules for three-point functions involving two 1/2 BPS operators}\label{selrul}

In this section we address the main subject of the present paper. The OPE of
two 1/2 BPS operators is determined by the three-point functions of the
following type:
\begin{equation}\label{3ptf}
  \langle \mbox{BPS}_{1/2}^{(m)}(1)\; \mbox{BPS}_{1/2}^{(n)}(2)\; {\cal O}^{(jk)}_{\ell
  s}(3)\rangle\,.
\end{equation}
Here $\mbox{BPS}_{1/2}^{(m)}$ and $\mbox{BPS}_{1/2}^{(n)}$  denote two 1/2 BPS
short operators described in Section \ref{compoBPS}. The third operator in eq.
(\ref{3ptf}) is characterized by the quantum numbers of its lowest
($\theta_3=0$) component (``superconformal primary field"). These are:
conformal dimension $\ell$ (a priori arbitrary), Lorentz spin $s$ (meaning that
the component is a symmetric traceless rank $s$ tensor) and an R symmetry irrep
labeled by a pair of integers $(jk)$. The latter appears in the tensor product
$(m)\otimes(n)$ of two of the irreps listed in (\ref{1/2BPSm}) (we assume that
$m\geq n$):
\begin{equation}\label{tensprod}
 (m)\otimes(n) =  \left\{
    \begin{array}{ll}
    D=3: & \bigoplus_{k=0}^{n}\bigoplus_{j=0}^{n-k}[0,j,m+n-2k-2j,0] \\
    D=4: & \bigoplus_{k=0}^{n}\bigoplus_{j=0}^{n-k}[j,m+n-2k-2j,j] \\
    D=5: & \bigoplus_{k=0}^{n}[m+n-2k] \\
    D=6: & \bigoplus_{k=0}^{n}\bigoplus_{j=0}^{n-k}[2j,m+n-2k-2j]
  \end{array}
  \right.
\end{equation}

In what follows we show that the few rather elementary facts about
supersingletons and their products we have presented in Section 2 are
sufficient to explain the selection rules on the operator ${\cal
O}^{(jk)}_{\ell s}$ in (\ref{3ptf}). Although we will be discussing three-point
functions {\it in superspace}, we will hardly need to go into any details of
their $\theta$ dependence. The examination of the lowest
($\theta_1=\theta_2=\theta_3=0$) component will give us all the necessary
information. The reason for this is the remarkable property of the three-point
functions (\ref{3ptf}) that they are uniquely fixed by conformal supersymmetry.
Indeed, the superfunction (\ref{3ptf}) depends on half of the Grassmann
variables at points 1 and 2 and on a full set of such variables at point 3.
Thus, the total number of odd variables exactly matches the number of
supersymmetries (Poincar\'e $Q$ plus special conformal $S$). Therefore there
exist no nilpotent superconformal invariants made out of the $\theta$s and the
complete $\theta_{1,2,3}$ expansion of (\ref{3ptf}) is determined by its lowest
component. The latter is the three-point function of two scalars and one tensor
field, and is fixed by conformal invariance up to an overall factor.

Before proceeding, we would like to compare the method we follow here with earlier approaches \cite{AES,ES,HH,EFS,HH2}. There the origin of
the selection rules was related to the requirement of absence of harmonic
singularities (harmonic analyticity) at the higher levels of the $\theta$
expansion of the three-point function. This is certainly an equivalent
explanation, however here we insist upon the fact that harmonic analyticity is
nothing but the coordinate expression of R symmetry irreducibility. Thus, by
just analyzing the occurrence of the different R symmetry irreps in conjunction
with our knowledge of the supermultiplet structure, we are able to derive the
same selection rules without inspecting the actual harmonic or Grassmann
coordinate dependence.

\subsection{Factorization of the three-point functions}

The crucial observation is that the lowest component of the three-point
function (\ref{3ptf}) can be factorized as follows:
$$\langle \mbox{BPS}_{1/2}^{(m)}(1)\; \mbox{BPS}_{1/2}^{(n)}(2)\; {\cal O}^{(jk)}_{\ell
  s}(3)\rangle_{\theta_{1,2,3}=0} =$$
 \vspace{5mm}
\begin{equation}\label{figure}
  \begin{picture}(0,5000)
  \drawline\fermion[\S\REG](-15000,3000)[6000]
  \global\Xone=\pfrontx
  \global\Yone=\pfronty
  \global\Ytwo=\pbacky
  \global\advance\Ytwo by -1100
  \global\advance\Yone by 900
  \global\advance\Xone by -500
  \put(\Xone,\Ytwo){$(n)$}
  \put(\Xone,\Yone){$(m)$}
  \drawline\fermion[\SE\REG](\fermionfrontx,\fermionfronty)[4243]
  \global\Xthree=\pbackx
  \global\Ythree=\pbacky
  \drawline\fermion[\SW\REG](\fermionbackx,\fermionbacky)[4243]
  \global\advance\Xthree by 900
  \put(\Xthree,\Ythree){${\cal O}^{(jk)}_{\ell s}$}
  \global\advance\Xthree by 5000
  \put(\Xthree,\Ythree){$=$}
  \drawline\fermion[\S\REG](-2000,3000)[6000]
  \global\Xone=\pfrontx
  \global\Yone=\pfronty
  \global\Ytwo=\pbacky
  \global\advance\Ytwo by -1100
  \global\advance\Yone by 900
  \global\advance\Xone by -500
  \put(\Xone,\Ytwo){$(k)$}
  \put(\Xone,\Yone){$(k)$}
  \drawline\fermion[\SE\REG](\fermionfrontx,\fermionfronty)[4243]
  \global\Xthree=\pbackx
  \global\Ythree=\pbacky
  \drawline\fermion[\SW\REG](\fermionbackx,\fermionbacky)[4243]
  \global\advance\Xthree by 900
  \put(\Xthree,\Ythree){${\cal O}_{d s}$}
  \global\advance\Xthree by 4000
  \put(\Xthree,\Ythree){$\otimes$}
  \phantom{\drawline\fermion[\S\REG](10000,3000)[6000]}
  \global\Xone=\pfrontx
  \global\Yone=\pfronty
  \global\Ytwo=\pbacky
  \global\advance\Ytwo by -1500
  \global\advance\Yone by 900
  \global\advance\Xone by -1500
  \put(\Xone,\Ytwo){$(m-k)$}
  \put(\Xone,\Yone){$(n-k)$}
  \drawline\fermion[\SE\REG](\fermionfrontx,\fermionfronty)[4243]
  \global\Xthree=\pbackx
  \global\Ythree=\pbacky
  \drawline\fermion[\SW\REG](\fermionbackx,\fermionbacky)[4243]
  \global\advance\Xthree by 900
  \put(\Xthree,\Ythree){$(jk)$}
  \end{picture}
  \end{equation}

 \vspace{20mm}
 where  the new ``fake operator"  ${\cal
O}_{d s}$ is an R symmetry singlet, but it carries spin $s$ and dimension
\begin{equation}\label{tildeell}
  d = \ell - \frac{D-2}{2}(m+n-2k) \; .
\end{equation}

The two factors in the r.h.s. of (\ref{figure}) have the following structure.
The factor carrying the spin at point 3 is made out of the 3-point conformal
vector
\begin{equation}\label{5.8}
  Y^\mu = \frac{x_{13}^\mu}{x_{13}^2} - \frac{x_{23}^\mu}{x_{23}^2}
\end{equation}
and of the supersingleton two-point function $\langle W(1) W(2)
\rangle_{\theta_{1,2}=0}\ $. The latter is completely determined by just R
symmetry, translation and dilatation invariance  and is given by
\begin{equation}\label{2pt}
  \langle W(1) W(2) \rangle_{\theta_{1,2}=0} =
  \frac{(12)}{(x^2_{12})^{\frac{D-2}{2}}} \;.
\end{equation}
Here $(12)$ symbolizes the irreducible harmonic structure which carries the
quantum numbers of a HWS of the R symmetry group corresponding to the basic
supersingleton:
\begin{equation}\label{(12)}
  (12)\quad \Leftrightarrow \quad  \left\{
  \begin{array}{ll}
    D=3: & (u_1)^{+(+)[+]}_a\; (u_2)^{+(+)[+]}_a \\
    D=4,6: &  (u_1)^1_i(u_1)^2_j\; \epsilon^{ijkl} \; (u_2)^1_k(u_2)^2_l \\
    D=5: & (u_1)^1_i\; \epsilon^{ij} \; (u_2)^1_j
  \end{array}
\right.
\end{equation}
Thus, the complete first factor  in the r.h.s. of (\ref{figure}) has the form
\begin{equation}\label{spinfactor}
\begin{picture}(0,5000)
  \drawline\fermion[\S\REG](-10000,3000)[6000]
  \global\Xone=\pfrontx
  \global\Yone=\pfronty
  \global\Ytwo=\pbacky
  \global\advance\Ytwo by -1100
  \global\advance\Yone by 900
  \global\advance\Xone by -500
  \put(\Xone,\Ytwo){$(k)$}
  \put(\Xone,\Yone){$(k)$}
  \drawline\fermion[\SE\REG](\fermionfrontx,\fermionfronty)[4243]
  \global\Xthree=\pbackx
  \global\Ythree=\pbacky
  \drawline\fermion[\SW\REG](\fermionbackx,\fermionbacky)[4243]
  \global\advance\Xthree by 900
  \put(\Xthree,\Ythree){${\cal O}_{d s}$}
  \global\advance\Xthree by 4000
  \put(\Xthree,\Ythree){$=$}
  \end{picture}
  \left(\frac{(12)}{(x^2_{12})^{\frac{D-2}{2}}} \right)^k \ (Y^2)^{\frac{d-s}{2}}\
   Y^{\{\mu_1}\cdots Y^{\mu_s\}} \ .
\end{equation}
\vskip12mm

The second factor  in the r.h.s. of (\ref{figure}) accounts for the R symmetry
quantum numbers at point 3. It is entirely made out of supersingleton two-point
functions:
\begin{center}
  \begin{picture}(0,5000)
  \phantom{\drawline\fermion[\S\REG](-3500,2700)[6000]}
  \global\Xone=\pfrontx
  \global\Yone=\pfronty
  \global\Ytwo=\pbacky
  \global\advance\Ytwo by -1500
  \global\advance\Yone by 900
  \global\advance\Xone by -1500
  \put(\Xone,\Ytwo){$(m-k)$}
  \put(\Xone,\Yone){$(n-k)$}
  \drawline\fermion[\SE\REG](\fermionfrontx,\fermionfronty)[4243]
  \global\Xthree=\pbackx
  \global\Ythree=\pbacky
  \drawline\fermion[\SW\REG](\fermionbackx,\fermionbacky)[4243]
  \global\advance\Xthree by 900
  \put(\Xthree,\Ythree){$(jk)\qquad =$}
  \end{picture}
  \end{center}
  \vspace{12mm}
\begin{equation}\label{Rfactor}
  \left(\frac{(13)}{(x^2_{13})^{\frac{D-2}{2}}} \right)^{m-j-k}
  \left(\frac{(23)}{(x^2_{23})^{\frac{D-2}{2}}} \right)^{n-j-k}
  \left[\frac{(13)}{(x^2_{13})^{\frac{D-2}{2}}}\ \frac{(23^-)}{(x^2_{23})^{\frac{D-2}{2}}}
  - 1 \leftrightarrow 2 \right]^j  \ .
\end{equation}
In order to reproduce the structure of the R symmetry representations at point
3 listed in eq. (\ref{tensprod}) we have to use both types of supersingletons,
according to eq. (\ref{kssgl}). Thus, while the harmonic factors $(13)$, $(23)$
originate from two-point functions involving only HWS (recall (\ref{2pt})), the
factor $(23^-)$ (and similarly $(13^-)$) comes from a two-point function of the
mixed type, i.e., where the supersingleton at point 3 is not the HWS of the
corresponding representation:
\begin{equation}
  \langle W(2) W^-(3) \rangle_{\theta=0} =
  \frac{(23^-)}{(x^2_{23})^{\frac{D-2}{2}}} \;; \nonumber
\end{equation}
\begin{equation}\label{(13-)}
  (23^-)\quad \Leftrightarrow \quad  \left\{
  \begin{array}{ll}
    D=3: & (u_2)^{+(+)[+]}_a\; (u_3)^{+(+)[-]}_a \\
    D=4,6: &  (u_2)^1_i(u_2)^2_j\; \epsilon^{ijkl} \; (u_3)^1_k(u_3)^3_l
     \end{array}
\right.
\end{equation}
(no such factors are needed in the case $D=5$). The irreducibility conditions
at point 3, eqs. (\ref{D3cond}) or (\ref{D4cond}), are then automatically
satisfied, given the fact that the raising operators act on the harmonics as
follows:
\begin{equation}\label{raisharm}
  D^{[++]}u^{+(+)[-]}_a =  u^{+(+)[+]}_a\;, \qquad D^2_3 u^3_1 = u^2_1
\end{equation}
and the antisymmetry of the factor $[\cdots]^j$ in eq. (\ref{Rfactor}). (Note
that the first factor (\ref{spinfactor}), being an R symmetry singlet at point
3, does not depend on the harmonics at that point.)

\subsection{BPS shortness and selection rules}

The form of the lowest component of the three-point function that we found in
the preceding subsection satisfies the basic requirements of conformal
covariance and R symmetry irreducibility. The construction we presented clearly
shows that this form is unique (up to an overall constant factor). Next, we
have to extend this lowest component to a full superspace three-point function.
According to the counting argument from the beginning of this section, this
superextension is also unique (if it exists).

One way to proceed would be to use various techniques to construct the
superconformal three-point covariant starting from its first component
\cite{Park,Osborn,Park2,KuzTheis,HH2}. This results in rather complicated
expressions which are not so easy to analyze. Here we present a different
approach based on the factorized form (\ref{figure}) which directly leads to
the conditions for existence of such a superextension.

The origin of possible constraints on the three-point functions is in the fact
that the operators at points 1 and 2 are 1/2 BPS short. The second factor
(\ref{Rfactor}) from eq. (\ref{figure}) can immediately be extended to a
superfunction having the required properties at points 1 and 2. Indeed, each
factor in eq. (\ref{Rfactor}) is the lowest component of a two-point function
of supersingletons, so the obvious superextension of eq. (\ref{Rfactor}) is
\begin{center}
  \begin{picture}(0,5000)
  \phantom{\drawline\fermion[\S\REG](-3500,3500)[6000]}
  \global\Xone=\pfrontx
  \global\Yone=\pfronty
  \global\Ytwo=\pbacky
  \global\advance\Ytwo by -1500
  \global\advance\Yone by 900
  \global\advance\Xone by -1500
  \put(\Xone,\Ytwo){$(m-k)$}
  \put(\Xone,\Yone){$(n-k)$}
  \drawline\fermion[\SE\REG](\fermionfrontx,\fermionfronty)[4243]
  \global\Xthree=\pbackx
  \global\Ythree=\pbacky
  \drawline\fermion[\SW\REG](\fermionbackx,\fermionbacky)[4243]
  \global\advance\Xthree by 900
  \put(\Xthree,\Ythree){$(jk)\qquad \Longrightarrow$}
  \end{picture}
  \end{center}
  \vspace{10mm}
\begin{equation}\label{2ptfactor}
  \langle W(1)W(3)\rangle^{m-j-k}
  \langle W(2)W(3)\rangle^{n-j-k}
  \left[\langle W(1)W(3)\rangle\ \langle W(2)W^-(3)\rangle
  - 1\leftrightarrow2   \right]^j
\end{equation}
Here $W^-(3)$ denotes the alterative realization (\ref{altern}) of the
supersingleton.

As explained in Section \ref{singl}, BPS shortness is equivalent to
analyticity, i.e., it is a multiplicative property. The product
(\ref{2ptfactor}) of BPS objects automatically has the required properties at
points 1 and 2. So, we need to concentrate on the first factor
(\ref{spinfactor}). We can formally treat this factor as the lowest component
of another, ``fake" three-point function involving two 1/2 BPS short objects of
identical labels $(k)$ at points 1 and 2 and an R symmetry singlet, a priori
long multiplet at point 3. Such a three-point function is also  uniquely fixed
by conformal supersymmetry, if it exists. To find out under what conditions
this lowest component can be extended to a complete superfunction, let us
restrict our attention to point 1. Then we can view the expression
(\ref{spinfactor}) as the lowest component of a 1/2 BPS short composite of the
type $[W]^k$. Now, according to the discussion of Section \ref{singl}, there
are only two cases where this lowest component must satisfy constraints
following from the conditions of BPS shortness. These cases are:

{\it (i)} If $k=0$ the only possible 1/2 BPS operator is the identity, so the
three-point function (\ref{spinfactor}) must be trivial. This implies $s=0$ and
$d=0$ or equivalently,
\begin{equation}\label{k=0}
Case\ (i):\qquad  k=0, \quad s=0, \quad \ell =
  \frac{D-2}{2}(m+n)\;.
\end{equation}

{\it (ii)} If $k=1$ we are dealing with the lowest component of a
supersingleton, i.e., with a massless scalar at point 1. Consequently, we must
require
\begin{equation}\label{box}
  k=1 \ \Rightarrow \qquad \square_1\left[(x^2_{12})^{-\frac{D-2}{2}}\ (Y^2)^{\frac{d-s}{2}}\ Y^{\{\mu_1}\cdots
  Y^{\mu_s\}}  \right] = 0\;.
\end{equation}
This equation is conformally covariant by construction, which allows us to go
to a special coordinate frame where it becomes very simple. Multiplying it by
$(x^2_{23})^{\frac{D-2}{2}}$ (this does not affect the differential operator
$\square_1$) and using translations and conformal boosts to set $x_3=0$,
$x_2=\infty$, we obtain (denoting $x_1\equiv x$):
\begin{equation}\label{bbox}
  \square\left[(x^2)^{-\frac{d+s}{2}}\ x^{\{\mu_1}\cdots
  x^{\mu_s\}}  \right] = 0\;.
\end{equation}
It is easy to see that this equation admits two solutions. The first solution
is $d=-s$ or equivalently,
\begin{equation}\label{Ist}
 \ell = -s + \frac{D-2}{2}(m+n-2)\,.
\end{equation}
In order for this solution to not violate the unitarity bounds on the
superconformal representation at point 3 of Section \ref{Semishort multiplets},
we must set $s=0$ which results in
\begin{equation}\label{BPSseries}
Case\ (ii.a):\qquad  k=1\;, \quad s=0\;, \quad  \ell = \frac{D-2}{2}(m+n-2)\,.
\end{equation}
The second solution to eq. (\ref{bbox}) is $d=s+D-2$ which gives rise to
\begin{equation}\label{IInd}
Case\ (ii.b):\qquad   k=1\;, \quad \ell = s + \frac{D-2}{2}(m+n)
\end{equation}
for arbitrary $s$.

We stress upon the fact that although the constraints above have been derived
as necessary  conditions, they are also sufficient for the existence of the
corresponding three-point functions. The argument is the same as stating that
given a massless scalar in the appropriate R symmetry representation, conformal
supersymmetry enables us to reconstruct the entire supersingleton multiplet
whose lowest component this scalar is. Similarly, when $k\geq2$ we are dealing
with the situation in which conformal supersymmetry restores the full 1/2 BPS
composite operator $[W]^k$ starting from an {\it unconstrained} scalar lowest
component. Hence, for $k\geq2$ we do not find any restrictions on the quantum
numbers $\ell,s$ at point 3, apart from the unitarity bounds from Section
(\ref{Semishort multiplets}). This is true even in the particular case $k=2$
when the bilinear $[W]^2$ is a supercurrent multiplet: The required conserved
vector (tensor) components will be automatically created as conformal
supersymmetry descendents of the lowest, unconstrained scalar component.

\section{Interpretation of the results. Protected operators}

In this section we show that the three cases (\ref{k=0}), (\ref{BPSseries}) and
(\ref{IInd}) correspond to {\it protected operators} at point 3, namely, to 1/2
or 1/4 BPS short operators in cases {\it (i)} and {\it (ii.a)} and to semishort
operators in case {\it (ii.b)}.

{\it Case (i)}

The simplest situation occurs when $k=0$. Then the factor (\ref{spinfactor}) in
eq. (\ref{figure}) becomes trivial and the entire three-point function is
reduced to the second factor (\ref{Rfactor}) which is just a product of
two-point functions of supersingletons, see eq. (\ref{2ptfactor}):
\begin{equation}\label{k=0BPS}
\langle \mbox{BPS}_{1/2}^{(m)}(1)\; \mbox{BPS}_{1/2}^{(n)}(2)\;
\mbox{BPS}(3)\rangle=
\end{equation}
$$
   \langle W(1)W(3)\rangle^{m-j}
  \langle W(2)W(3)\rangle^{n-j}
  \left[\langle W(1)W(3)\rangle\ \langle W(2)W^-(3)\rangle
  - 1\leftrightarrow2   \right]^j  \;.
$$
Note the absence of a two-point function connecting points 1 and 2. Now we can
identify the operator at point 3 with a composite BPS operator (recall eq.
(\ref{kssgl})):
\begin{equation}\label{BPScase}
\begin{array}{ll}
     \mbox{BPS}_{1/2}(3)=[W(3)]^{m+n} & \mbox{if $j=0$} \\
    \mbox{BPS}_{1/4}(3)=[W(3)]^{m+n-j}\; [W^-(3)]^{j} & \mbox{if $j\neq0$}
  \end{array}
\end{equation}

{\it Case (ii.a)}

In this case the three-point function still factorizes into two-point functions
of supersingletons:
\begin{equation}\label{k=1BPS}
\langle \mbox{BPS}_{1/2}^{(m)}(1)\; \mbox{BPS}_{1/2}^{(n)}(2)\;
\mbox{BPS}(3)\rangle=
\end{equation}
$$
 \langle W(1)W(2)\rangle \langle W(1)W(3)\rangle^{m-j-1}
  \langle W(2)W(3)\rangle^{n-j-1} \left[\langle W(1)W(3)\rangle\ \langle W(2)W^-(3)\rangle
  - 1\leftrightarrow2   \right]^j
$$
and the operator at point 3 is
\begin{equation}\label{BPScase2}
\begin{array}{ll}
    \mbox{BPS}_{1/2}(3)=[W(3)]^{m+n-2} & \mbox{if $j=0$} \\
    \mbox{BPS}_{1/4}(3)=[W(3)]^{m+n-2-j}\; [W^-(3)]^{j} & \mbox{if $j\neq0$}
  \end{array}
\end{equation}

{\it Case (ii.b)}

In this case the three-point function does not factorize into two-point
functions. Instead, we remark that the conformal dimension of the operator
${\cal O}^{(j1)}_{\ell s}$ at point 3 takes the special values listed in eq.
(\ref{specialell}) which characterize a ``semishort" operator. We can then
conjecture that the three-point function becomes
\begin{equation}\label{k=1semi}
\langle \mbox{BPS}_{1/2}^{(m)}(1)\; \mbox{BPS}_{1/2}^{(n)}(2)\; {\cal
S}^{\{\mu_1\cdots\mu_s\}\;(j1)}(3)\rangle\,.
\end{equation}

However, the fact that the operator ${\cal O}^{(j1)}_{\ell s}$ has the right
quantum numbers is not yet sufficient to claim that it is indeed semishort. A
simple counterexample illustrating this point is the three-point function of
two scalars of {\it different} dimensions $\ell_{1,2}$ and a vector of {\it
canonical} dimension $\ell_j= D-1$:
\begin{equation}\label{counterex}
  \langle \phi_{\ell_1}(1)  \phi_{\ell_2}(2) j^\mu(3)
   \rangle = (x^2_{12})^{\frac{D-2-\ell_1-\ell_2}{2}}
  (x^2_{13})^{\frac{\ell_2-\ell_1-D+2}{2}}(x^2_{23})^{\frac{\ell_1-\ell_2-D+2}{2}}
  \left( \frac{x_{13}^\mu}{x_{13}^2} - \frac{x_{23}^\mu}{x_{23}^2}  \right)\;.
\end{equation}
A direct calculation shows that the vector at point 3 is conserved if and only
if the two scalars have the {\it same dimension}, $\ell_1=\ell_2$. The same is
true if we replace the vector $j^\mu$ by any symmetric traceless tensor
$j^{\{\mu_1\cdots\mu_s\}}$ of canonical dimension $\ell=s+D-2$.

Thus, we need to provide additional evidence that the operator ${\cal
O}^{(j1)}_{\ell s}$ is semishort. In earlier publications \cite{ES,EFS} we have
done this by restoring the $\theta_3$ dependence of the three-point function
and then showing directly that the superspace constraints of the type
(\ref{conconproj}) are satisfied. Here we present a much simpler argument
based, once again, on the factorization (\ref{figure}) of the three-point
function, on the one hand, and on the composite form of the operators
(\ref{genersuperconf}), on the other hand. We already know that the operator
${\cal O}^{(j1)}_{\ell s}$ can be factorized as follows:
\begin{equation}\label{factform}
  {\cal O}^{(j1)}_{\ell s} = {\cal O}_{ds}\; \mbox{BPS}^{(j1)}
\end{equation}
where the BPS factor $\mbox{BPS}^{(j1)}$ is given in eq. (\ref{2ptfactor}).
Then we should expect that the remaining singlet factor ${\cal O}_{ds}$ can be
identified with a ``supercurrent" $J^{\{\mu_1\cdots\mu_s\}}$ if $s>0$, as in
(\ref{semispin}) (or with a scalar ``supercurrent" $\Phi$ if $s=0$, as in
(\ref{seminospin})). That this is the case is evident from the bosonic example
above. Indeed, now the first factor (\ref{spinfactor}) of the lowest component
(\ref{figure}) corresponds to the three-point function of two scalars of {\it
equal} dimensions $\ell_{1}=\ell_2=(D-2)/2$ and a tensor of {\it canonical}
dimension $\ell= s+D-2$. Consequently, this tensor must be conserved. Since it
is the first component\footnote{In the case $s=0$ the first conserved component
of the supercurrent $\Phi$ is not the lowest component of the supermultiplet,
but this does not affect the argument.} of a supermultiplet, its conservation
implies that the operator ${\cal O}_{ds}$ does satisfy a ``supercurrent"
constraint of the type (\ref{concon}),
$$
  {\cal O}_{ds} \ \rightarrow \ J^{\{\mu_1\cdots\mu_s\}}\,.
$$
Thus, we can interpret (\ref{spinfactor}) as the lowest component of a ``fake"
three-point function,
\begin{equation}\label{spinfactor2}
\begin{picture}(0,5000)
  \drawline\fermion[\S\REG](-10000,3000)[6000]
  \global\Xone=\pfrontx
  \global\Yone=\pfronty
  \global\Ytwo=\pbacky
  \global\advance\Ytwo by -1100
  \global\advance\Yone by 900
  \global\advance\Xone by -500
  \put(\Xone,\Ytwo){$(1)$}
  \put(\Xone,\Yone){$(1)$}
  \drawline\fermion[\SE\REG](\fermionfrontx,\fermionfronty)[4243]
  \global\Xthree=\pbackx
  \global\Ythree=\pbacky
  \drawline\fermion[\SW\REG](\fermionbackx,\fermionbacky)[4243]
  \global\advance\Xthree by 900
  \put(\Xthree,\Ythree){${\cal O}_{s+D-2,\; s}$}
  \global\advance\Xthree by 5500
  \put(\Xthree,\Ythree){$=$}
  \end{picture}
  \ \ \langle
\mbox{BPS}_{1/2}^{(1)}(1)\; \mbox{BPS}_{1/2}^{(1)}(2)\;
J^{\{\mu_1\cdots\mu_s\}}(3)\rangle_{\theta_{1,2,3}=0}\ .
\end{equation}
\vskip12mm
Finally, the presence of the factor $\mbox{BPS}^{(j1)}$ weakens the
constraint on the composite operator ${\cal O}^{(j1)}_{\ell s}$ and turns it
into a semishort operator
$$
  {\cal
O}^{(j1)}_{\ell s}\ \rightarrow \ {\cal S}^{\{\mu_1\cdots\mu_s\}\;(j1)}=
J^{\{\mu_1\cdots\mu_s\}}\; \mbox{BPS}^{(j1)}
$$
which proves (\ref{k=1semi}).

It is important to realize that the above factorization is only formal, it just
helps us investigate the supermultiplet structure. In fact, the singlet factor
${\cal O}_{ds}$ which looks like a ``supercurrent" is not the true operator at
point 3. The full operator ${\cal O}^{(j1)}_{\ell s}$ is only semishort and not
a supercurrent (and, consequently, does not contain conserved tensor
components). The reason is that in the case $k=1$ the BPS factor
$\mbox{BPS}^{(j1)}$ is always present. Indeed, from (\ref{tensprod}) it follows
that in order not to have a BPS factor the operator ${\cal O}^{(j1)}_{\ell s}$
must be an R symmetry singlet. Thus, we should set $j=0$ and $m+n=2$ which
implies $m=n=1$. However, this corresponds to putting just one supersingleton
at points 1 and 2 which is not a gauge invariant object and thus is not of
physical relevance.

The same argument shows that if $k\geq 2$ one can have a situation where the
BPS factor is absent and the operator ${\cal O}^{(jk)}_{\ell s}$ is a true
``supercurrent". Going back to the generic factorized form
(\ref{genersuperconf}), we see that this may happen if $j=0$, $m+n=2k\geq4$ and
if we {\it choose to set} $s>0$, $\delta=0$ or $s=0$, $\delta=1$. A well-known
example is the  Konishi multiplet in $N=4$, $D=4$ SYM which appears in the OPE of two stress-tensor multiplets and corresponds
to $m=n=k=2$, $j=s=0$. In the free field theory it is known to satisfy a superspace
constraint and to contain conserved tensor components. However, in the presence of interactions the Konishi multiplet acquires an anomalous dimension \cite{Anselmi:1997dd} and thus seizes to be a ``supercurrent". Further examples of operators which have anomalous dimension are the higher-spin and R symmetry singlet multiplets ($m=n=k=2$, $j=0$ and $s>0$) considered in \cite{Anselmi:2000bb}. These operators again reduce to ``supercurrents" in the free field theory.

The above discussion clearly shows the key difference between the cases $k=0,1$
and $k\geq2$. In the cases $k=0,1$ the conformal dimension at point 3 is fixed
by the branching rules and thus ${\cal O}$ {\it necessarily} becomes BPS or
semishort. In the case $k\geq 2$ there is no reason to maintain the conformal
dimension at one of these fixed values, so ${\cal O}$ may be a BPS, a semishort
or a generic long multiplet. It follows that for $k=0,1$ any operator ${\cal
O}^{(jk)}_{\ell s}$ appearing in the OPE of two 1/2 BPS operators is {\it
protected} by the superconformal kinematics whereas for $k\geq2$ it is {\it
unprotected}, i.e., its conformal dimension is determined by the dynamics of
the theory.

\section{Extremal correlators}

One of the possible applications of the branching rules that we have found is
the proof that a certain class of $n$--point correlation functions of 1/2 BPS
operators ${\cal W}^m \equiv [W]^m$,
\begin{equation}\label{e1}
  \langle {\cal W}^{m_1}(1) {\cal W}^{m_2}(2) \cdots {\cal W}^{m_n}(n)
  \rangle\;,
\end{equation}
remain non-renormalized in the interacting
theory. According to the terminology introduced in \cite{DHoFrMaMaRa} they are
called ``extremal" if
\begin{equation}\label{e2}
  m_1 =\sum_{i=2}^n m_i \,.
\end{equation}
Using AdS supergravity considerations, in \cite{DHoFrMaMaRa,DP} it was shown that
the extremal correlators are not renormalized and factorize into products of
two-point functions. Here we give a simple explanation of this fact from the
CFT point of view based on our results on the three-point functions and the
related OPEs of 1/2 BPS operators. The same argument has already been presented
in \cite{EFS} for the case $D=6$.\footnote{A different proof in the case $D=4$, based on a direct analysis of the $n$-point superconformal covariants, was given in \cite{Eden:2000gg}. See also \cite{HH2} for a recent argument.}

For simplicity we restrict ourselves to four-point extremal correlators. They
can be represented as the convolution of two OPEs:
\begin{eqnarray}
  && \langle {\cal W}^{m_1}(1) {\cal W}^{m_2}(2) {\cal W}^{m_3}(3) {\cal W}^{m_4}(4) \rangle =
  \label{e3} \\
  &&\qquad \sum \int_{5,5'}
  \langle {\cal W}^{m_1}(1) {\cal W}^{m_2}(2) {\cal O}(5) \rangle \
  \langle  {\cal O}(5)  {\cal O}(5') \rangle^{-1} \
   \langle  {\cal O}(5') {\cal W}^{m_3}(3) {\cal W}^{m_4}(4)\rangle \nonumber
\end{eqnarray}
where the sum goes over all possible operators which appear in the intersection
of the two OPEs. Due to the orthogonality of different operators the inverse
two-point function $\langle  {\cal O}(5)  {\cal O}(5') \rangle^{-1}$ only
exists if ${\cal O}(5)$ and ${\cal O}(5')$ are identical. To find out their
spectrum, we first examine the R symmetry quantum numbers. From
(\ref{tensprod}) we see that they are given by a pair of integers:
\begin{eqnarray}
 {\cal O}(5)\,: &&
 j,\ m_1+m_2-2j-2k\, ,\qquad 0\leq k\leq m_2, \ 0\leq j\leq m_2-k \nonumber\\
 {\cal O}(5')\,: &&
 j',\ m_3+m_4-2j'-2k'\,, \qquad 0\leq k'\leq m_4, \ 0\leq j'\leq m_4-k' \label{e4}
\end{eqnarray}
where we have assumed $m_3 \geq m_4$. Since in the extremal case
$m_1=m_2+m_3+m_4$ (recall (\ref{e2})), the intersection is given by the
following conditions:
\begin{equation}\label{e5}
  j=j'\,, \qquad 0 \leq k' = k-m_2 \leq 0\,,
\end{equation}
whose only solution is
\begin{equation}\label{e6}
  k=m_2 \ \Rightarrow \ j=j'=0\,, \qquad k'=0 \,.
\end{equation}
Further, from (\ref{BPScase}) we deduce that $k'=0$ and $j'=0$ imply that
${\cal O}(5')$, and by orthogonality, ${\cal O}(5)$ must be identical 1/2 BPS
operators,
\begin{equation}\label{e7}
  {\cal O} =  {\cal W}^{m_3+m_4}\,.
\end{equation}
Finally, in this particular case the three-point functions in (\ref{e3})
degenerate into products of two two-point functions (recall (\ref{k=0BPS})), so
(\ref{e3}) becomes
\begin{eqnarray}
  && \hskip-2cm \langle {\cal W}^{m_1}(1) {\cal W}^{m_2}(2) {\cal W}^{m_3}(3) {\cal W}^{m_4}(4) \rangle
  \label{e8} \\
  &=&\int_{5'}  \langle W(1) W(2) \rangle^{m_2} \ \int_{5} \langle W(1) W(5) \rangle^{m_3+m_4} \
  \langle W(5) W(5') \rangle^{-(m_3+m_4)}    \nonumber\\
  &&\qquad \times  \langle W(5') W(3) \rangle^{m_3}\
   \langle W(5') W(4) \rangle^{m_4}      \nonumber\\
  &=& \langle W(1) W(2) \rangle^{m_2} \ \langle W(1) W(3) \rangle^{m_3} \
  \langle W(1) W(4) \rangle^{m_4} \,. \nonumber
\end{eqnarray}
This clearly shows that the extremal four-point correlator factorizes into a
product of two-point functions. In other words, it always takes its free (Born
approximation) form, so it stays {\it non-renormalized}.

The generalization of the above result to an arbitrary number of points is
straightforward and it follows the $D=6$ pattern exhibited in \cite{EFS}.
Further, the argument concerning the non-renormalization of ``next-to-extremal"
\cite{DHoFrMaMaRa1,D'Hoker:2000dm} $D=6$ correlators (i.e., those for which
$m_1 =\sum_{i=2}^n m_i - 2$) presented in \cite{EFS} applies to the cases
$D=3,4,5$ as well.

Notice that three-point functions of protected operators in $D=4$, other than
1/2 BPS, have recently been proved not to suffer from renormalization
\cite{HH2,nonren3pt}.

\section*{Acknowledgements}
The work of S.F. has been supported in part by the European Commission RTN
network HPRN-CT-2000-00131 (Laboratori Nazionali di Frascati, INFN) and by the
D.O.E. grant DE-FG03-91ER40662, Task C. E.S. acknowledges numerous discussions
with Burkhard Eden.


\begin{thebibliography}{99}

\bibitem{AGMOO}
O. Aharony, S.S. Gubser, J. Maldacena, H. Ooguri and Y. Oz, { Phys. Rept.}
{\bf 323} (2000) 183,  hep-th/9905111.


\bibitem{Frrev}
D.Z. Freedman and P. Henry-Labordere, ``Field theory insight from the AdS/CFT
correspondence,'' hep-th/0011086.

\bibitem{Brev} M.~Bianchi,
``(Non-)perturbative tests of the AdS/CFT correspondence,'' hep-th/0103112.

\bibitem{HWrev}
P.~S.~Howe and P.~C.~West,
Class. Quant. Grav. {\bf 18} (2001) 3143, hep-th/0105218.

\bibitem{Haag:1975qh}
R.~Haag, J.~T.~Lopuszanski and M.~Sohnius,
Nucl. Phys. {\bf B88} (1975) 257.

\bibitem{Nahm:1978tg}
W.~Nahm,
Nucl. Phys. {\bf B135} (1978) 149.

\bibitem{AFSZ}
L. Andrianopoli, S. Ferrara, E. Sokatchev and B. Zupnik, { Adv. Theor. Math.
Phys.} \textbf{3} (1999) 1149, hep-th/9912007.

\bibitem{FS1}
S. Ferrara and E. Sokatchev, { Int. J. Theor. Phys.} {\bf 40} (2001) 935,
hep-th/0005151.

\bibitem{GIK1}
A. Galperin, E. Ivanov, S. Kalitzin, V. Ogievetsky and E. Sokatchev, Class.
Quant. Grav. {\bf 1} (1984) 469.

\bibitem{TheBook}
A. Galperin, E. Ivanov, V. Ogievetsky and E. Sokatchev, {\em Harmonic
superspace}, CUP, 2001.

\bibitem{GIO}
A. Galperin, E. Ivanov and V.I. Ogievetsky, { JETP Lett.} {\bf 33} (1981)
168.

\bibitem{hw1}
P.S. Howe and P.C. West, \ijmp14(1999)2659, \xxx9509140; \plb400(1997)307,
\xxx9611075; {\em Is $N=4$ Yang-Mills soluble?} (1996) in Moscow 1996,
\xxx9611074.

\bibitem{Ferrara:1998ej}
S.~Ferrara, C.~Fronsdal and A.~Zaffaroni,
Nucl. Phys.  {\bf B532} (1998) 153, hep-th/9802203.

\bibitem{AF} L. Andrianopoli and S. Ferrara,
{\bf Phys. Lett.} B430 (1998) 248, hep-th/9803171;
 {\bf Lett. Math. Phys.} 46 (1998) 265, hep-th/9807150;
 {\bf Lett. Math. Phys.} 48
(1999) 145, hep-th/9812067.

\bibitem{RomNieu} L.J. Romans and P. van Nieuwenhuizen,
{ Phys. Rev.} {\bf D32} (1985) 389.

\bibitem{gm}
M. G\"unaydin and N. Marcus, { Class. Quantum Grav.} {\bf 2} (1985) L11.

\bibitem{AFP1}
G. Arutyunov, S. Frolov and A.C. Petkou, \npb586(2000)547, \xxx0005182.

\bibitem{AFP2}
G. Arutyunov, S. Frolov and A.C. Petkou, \npb602(2001)238, \xxx0010137.

\bibitem{AEPS}
G. Arutyunov, B. Eden, A.C. Petkou and E. Sokatchev, { ``Exceptional
non-renormalization properties and OPE analysis of chiral  four-point functions
in N = 4 SYM(4)"}, \xxx0103230.


\bibitem{AES}
G. Arutyunov, B. Eden, E. Sokatchev, ``On non-renormalization and OPE in
superconformal field theories'', hep-th/0105254.

\bibitem{BKRSkonishi}
M. Bianchi, S. Kovacs, G. Rossi and Y.S. Stanev, \jhep0105(2001)042,
\xxx0104016.

\bibitem{FrGuWa} D.Z. Freedman, S.S. Gubser, K. Pilch and N.P. Warner, {
Adv. Theor. Math. Phys.} {\bf 3} (1999) 363, hep-th/9904017.

\bibitem{ES}
B. Eden and E. Sokatchev, ``On the OPE of 1/2 BPS short operators in $N=4$
SCFT$_4$'', hep-th/0106249.

\bibitem{HH}
P.J. Heslop and P.S. Howe, {
Phys. Lett.} {\bf B516} (2001) 367,
hep-th/0106238.

\bibitem{Rom} L.J. Romans, { Phys. Lett.} {\bf B153} (1985) 392.

\bibitem{CerFer} A. Ceresole, G. Dall'Agata, R. D'Auria and S. Ferrara,
{ Phys. Rev.} {\bf D61} (2000) 066001, hep-th/9905226.

\bibitem{KleWit} I. Klebanov and E. Witten, { Nucl. Phys.}
{\bf B536} (1998) 199, hep-th/9807080.

\bibitem{DHoFrMaMaRa}
E. D'Hoker, D.Z. Freedman, S.D. Mathur, A. Matusis and L. Rastelli, ``Extremal
correlators in the AdS/CFT correspondence'', In Shifman, M.A. (ed.): {\it The many faces of the superworld} 332-360, \xxx9908160.

\bibitem{DP}
E. D'Hoker and B. Pioline, { JHEP} {\bf 0007} (2000) 021, hep-th/0006103.



\bibitem{HH2}
P.~J.~Heslop and P.~S.~Howe,
{``OPEs and 3-point correlators of protected operators in N = 4 SYM"},
hep-th/0107212.


\bibitem{EFS}
B.~Eden, S.~Ferrara and E.~Sokatchev,
{``(2,0) superconformal OPEs in D = 6, selection rules and  non-renormalization theorems"}, hep-th/0107084.


\bibitem{Ferrara:2000zg}
S.~Ferrara and E.~Sokatchev,
Lett. Math. Phys.  {\bf 52} (2000) 247, hep-th/9912168.


\bibitem{FS2}
S. Ferrara and E. Sokatchev, { Lett. Math. Phys.} {\bf 51} (2000) 55,
hep-th/0001178.

\bibitem{HHowe}
P. Heslop and P.S. Howe, { Class. Quant. Grav.}  {\bf 17} (2000) 3743, hep-th/0005135.

\bibitem{Fron} C. Fronsdal, { Lett. Math. Phys.} {\bf 16} (1988) 163, 173.

\bibitem{Duf} M.J. Duff, { Int. J. Mod. Phys.} {\bf A14} (1999) 815, hep-th/9808100.

\bibitem{Nicolai:1984gb}
H.~Nicolai and E.~Sezgin,
{ Phys. Lett.} {\bf  B143} (1984) 389.

\bibitem{Gunaydin:1985wc}
M.~Gunaydin, P.~van Nieuwenhuizen and N.~P.~Warner, \npb255(1985)63.

\bibitem{FT}  S. Ferrara and M. Testa, { Phys. Lett.} \textbf{B49} (1974) 95.

\bibitem{FP}
E.S. Fradkin and M.Ya. Palchik, \rep44(1978)249; {\em Conformal Quantum Field
Theory in D-Dimensions}, Kluwer (1996) 461p.

\bibitem{Siegel2} W.~Siegel,
Int. J. Mod. Phys. {\bf A4} (1989) 2015.

\bibitem{EHWa} T. Enright, R. Howe and N. Wasllach, ``A classification of unitary highest weight modules", in {\it Representation Theory of Reductive Groups}, ed. P.C. Trombi, Birkh\"auser, 1982.


\bibitem{FeFo}
S. Ferrara and C. Fronsdal, ``Conformal fields in higher dimensions'',  presented at the International Conference on Quantization, Gauge Theory, and Strings: Conference Dedicated to the Memory of E. Fradkin, Moscow, Russia (2000), hep-th/0006009.

\bibitem{Siegel}
W. Siegel, Nucl. Phys. \textbf{B177} (1981) 325.

\bibitem{HST} P.S. Howe, K.S. Stelle and P.K. Townsend, { Nucl.
Phys.} {\bf B191} (1981) 445; { Nucl. Phys. } {\bf B192} (1981) 332.

\bibitem{Howe:1983fr}
P.~S.~Howe, G.~Sierra and P.~K.~Townsend,
\npb221(1983)331.

\bibitem{Parkkk} J.~H.~Park,
J. Math. Phys.  {\bf 41} (2000) 7129, hep-th/9910199.

\bibitem{Howe} P.S. Howe, in {\it Supersymmetries and Quantum
Symmetries}, eds. J. Wess and E.A. Ivanov, Lecture Notes in Physics, Springer
(1999), p. 68-78,  hep-th/9812133.


\bibitem{hh}
G.G. Hartwell and P.S. Howe, Int. J. Mod. Phys. \textbf{A10} (1995)
3901, \xxx9412147; Class. Quant. Grav. \textbf{12} (1995) 1823.

\bibitem{Skiba:1999im}
W.~Skiba,
{ Phys. Rev.}  {\bf D60} (1999) 105038, hep-th/9907088.

\bibitem{FF}
M. Flato and C. Fronsdal, Lett. Math. Phys. {\bf 8} (1984) 159.

\bibitem{dp}
V. K. Dobrev and V. B. Petkova, Phys. Lett. {\bf B162} (1985) 127; Fortschr.
Phys. {\bf 35} (1987) 7, 537.

\bibitem{Minw2} S. Minwalla,
Adv. Theor. Math. Phys. {\bf 2} (1998) 781.

\bibitem{Park}
J.~H.~Park, \ijmp13(1998)1743, \xxx9703191.

\bibitem{Osborn}
H. Osborn, \ap272(1999)243, \xxx9808041.

\bibitem{Park2}
J.~H.~Park, Nucl. Phys. {\bf B559} (1999) 455, hep-th/9903230.

\bibitem{KuzTheis}
S.M. Kuzenko and S. Theisen, Class. Quant. Grav. {\bf 17} (2000) 665,
hep-th/9907107.

\bibitem{Anselmi:1997dd}
D.~Anselmi, D.~Z.~Freedman, M.~T.~Grisaru and A.~A.~Johansen,
Phys. Lett. {\bf B394} (1997) 329, hep-th/9608125.

\bibitem{Anselmi:2000bb}
D.~Anselmi,
Nucl. Phys. {\bf B541} (1999) 369,
hep-th/9809192;
Class. Quant. Grav.  {\bf 17} (2000) 1383, hep-th/9906167.


\bibitem{Eden:2000gg}
B.~U.~Eden, P.~S.~Howe, E.~Sokatchev and P.~C.~West,
Phys. Lett. {\bf B494} (2000) 141, hep-th/0004102; B.~Eden, P.~S.~Howe, C.~Schubert, E.~Sokatchev and P.~C.~West,
Phys. Lett.  {\bf B472} (2000) 323,  hep-th/9910150.


\bibitem{DHoFrMaMaRa1}
B.U. Eden, P.S. Howe, C. Schubert, E. Sokatchev and P.C. West,
\plb472(2000)323, \xxx9910150.

\bibitem{D'Hoker:2000dm}
E. D'Hoker, J. Erdmenger, D.Z. Freedman and M. Perez-Victoria,
Nucl. Phys. {\bf B589} (2000) 3, hep-th/0003218.

\bibitem{nonren3pt} E.~D'Hoker and A.~V.~Ryzhov,
``Three-point functions of quarter BPS operators in N = 4 SYM,''
hep-th/0109065.






\end{thebibliography}
\end{document}